\documentclass[twocolumn, showpacs, preprintnumbers, amsmath, amssymb]{revtex4}

\usepackage{graphicx}
\usepackage{dcolumn}
\usepackage{bm}
\usepackage{mathbbol}
\usepackage{axodraw}
\usepackage{epsf}
\usepackage{mathrsfs}
\usepackage{slashed}
\usepackage{stmaryrd}

\newcommand{\be}{\begin{equation}}
\newcommand{\ee}{\end{equation}}
\newcommand{\bea}{\begin{eqnarray}}
\newcommand{\eea}{\end{eqnarray}}
\newcommand{\bml}{\begin{mathletters} \baselineskip 10pt}
\newcommand{\eml}{\baselineskip 12pt \end{mathletters}}
\newcommand{\nn}{\nonumber}

\newcommand{\m}{{\scriptscriptstyle -}}
\newcommand{\p}{{\scriptscriptstyle +}}

\newcommand{\bra}{\langle}
\newcommand{\ket}{\rangle}

\newcommand{\dkp}{\frac{d^{d-2}k_\perp}{(2\pi)^{d-2}}}
\newcommand{\dlp}{\frac{d^{d-2}\ell_\perp}{(2\pi)^{d-2}}}

\def\lambdabar{\protect\@lambdabar}
\def\@lambdabar{%
\relax \bgroup
\def\@tempa{\hbox{\raise.73\ht0
\hbox to0pt{\kern.2\wd0\vrule width.7\wd0
height.1pt depth.1pt\hss}\box0}}%
\mathchoice{\setbox0\hbox{$\displaystyle\lambda$}\@tempa}%
{\setbox0\hbox{$\textstyle\lambda$}\@tempa}%
{\setbox0\hbox{$\scriptstyle\lambda$}\@tempa}%
{\setbox0\hbox{$\scriptscriptstyle\lambda$}\@tempa}%
\egroup }

\newcommand{\sfrac}[2]{{\textstyle \frac{#1}{#2}}}

\newcommand{\vc}[1]{\mbox{\boldmath$#1$}}

\newcommand{\ssvc}[1]{\mbox{\scriptsize\boldmath$#1$}}

\newcommand{\sgn}{{\mbox{sgn}}}

\begin{document}

\title{A novel approach to light-front perturbation theory}


\author{Thomas Heinzl}
\email{theinzl@plymouth.ac.uk}

\affiliation{School of Mathematics and Statistics, University of
Plymouth\\
Drake Circus, Plymouth PL4 8AA, UK}

\date{\today}

\begin{abstract}
We suggest a possible algorithm to calculate one-loop $n$-point
functions within a variant of light-front perturbation theory. The
key ingredients are the covariant Passarino-Veltman scheme and a
surprising integration formula that localises Feynman integrals at
vanishing longitudinal momentum. The resulting expressions are
generalisations of Weinberg's infinite-momentum results and are
manifestly Lorentz invariant. For $n$ = 2 and 3 we explicitly show
how to relate those to light-front integrals with standard energy
denominators. All expressions are rendered finite by means of
transverse dimensional regularisation.
\end{abstract}



\maketitle

\section{\label{sec:1}Introduction}

With precision tests of the Standard Model being routinely
performed nowadays perturbation theory based on covariant Feynman
rules has acquired a rather mature state. Initiated by the work of
Brown and Feynman \cite{brown:1952} and later `t~Hooft and Veltman
\cite{thooft:1979a} as well as Passarino and Veltman
\cite{passarino:1979} there is now a well established algorithm
(sometimes called Passarino-Veltman reduction) expressing
arbitrary one-loop $n$-point functions in terms of known basic
integrals (see \cite{bardin:1999} and \cite{denner:1991} for a
recent text and review, respectively).

The achievements of covariant perturbation theory continue to be
impressive but this seems far from also being true for
noncovariant approaches (at least within the realm of relativistic
quantum field theory). `Old-fashioned perturbation theory' based
on Hamiltonians defined at some given instant in time $t$ deserves
its name and is basically no longer in use (apart from
nonrelativistic applications e.g.\ in solid-state and many-body
physics). Instead of a single covariant $n$-point diagram one has
to evaluate $n!$ diagrams corresponding to the $n!$ time-orderings
of the $n$ vertices. This hugely increased effort is invested only
to find that the $n!$ noncovariant energy denominators
(resolvents) in Feynman integrands add up to the covariant answer,
the prototype relation being
\be \label{P0POLES}
  \frac{1}{p^2 - m^2} = \frac{1}{2E_p} \left( \frac{1}{p^0 - E_p} -
  \frac{1}{p^0 + E_p} \right) \;,
\ee
with the usual on-shell energy $E_p = (\vc{p}^2 + m^2)^{1/2}$.
Interpreting (\ref{P0POLES}) as a Feynman integrand the associated
diagram is a scalar tadpole which measures the (infinite) volume
of the mass shell.

Obviously, the factorial growth in the number of diagrams severely
obstructs feasible applications of Hamiltonian perturbation
theory. Only under special circumstances where most of the
noncovariant diagrams vanish might such an approach be considered.
Somewhat surprisingly, it turns out that such circumstances do in
fact exist. The crucial observation goes back to Weinberg
\cite{weinberg:1966a} who noted that many diagrams of
old-fashioned perturbation theory vanish in the
\textit{infinite-momentum limit}. Soon afterwards it was noted
\cite{chang:1969b,kogut:1970} that this limit coincides with
Hamiltonian perturbation theory based on Dirac's \textit{front
form} of relativistic dynamics \cite{dirac:1949}. In this approach
the dynamical evolution parameter is chosen to be light-front
time, $x^\p \equiv t + z$, rather than ordinary `Galileian' time
$t$ (Dirac's \textit{instant form}). Accordingly, the generator of
time evolution is the light-front Hamiltonian, $P^\m \equiv P^0 -
P^3$, describing time development from an initial light-front (or
null plane), say $x^\p = 0$, as light-front time goes by. Note
that null planes are somewhat peculiar from a Euclidean point of
view. Being tangential to the light-cone they contain light-like
directions perpendicular to $t$ and $z$ axes. In addition they
have light-like normals lying within the plane
\cite{rohrlich:1971}.

Given a 4-vector $a^\mu$ one introduces light-front coordinates
according to
\be
  a^\pm \equiv a^0 \pm a^3 \; , \quad \vc{a}_\perp \equiv (a^1 ,
  a^2) \; ,
\ee
such that the Minkowski scalar product becomes
\be
  a \cdot b = \sfrac{1}{2} a^\p b^\m  + \sfrac{1}{2} a^\m b^\p  -
  \vc{a}_\perp \cdot \vc{b}_\perp \; ,
\ee
which is \textit{linear} in both plus and minus components. This
has far reaching consequences. The on-shell light-front energy
becomes
\be \label{PMIN}
  [p^\m] \equiv \frac{\vc{p}_\perp^2 + m^2}{p^\p} \; ,
\ee
with positive and negative mass shell corresponding to positive
and negative longitudinal momentum, $p^\p$, separated by a pole at
$p^\p = 0$ rather than a mass gap of size 2$m$. Thus, instead of
(\ref{P0POLES}) one has a single `noncovariant' contribution only,
\be \label{PMINPOLE}
  \frac{1}{p^2 - m^2} = \frac{1}{p^\p} \; \frac{1}{p^\m - [p^\m]} \; .
\ee
This is a slightly simplistic version of Weinberg's original
observation \cite{weinberg:1966a} which since then has evolved
into an alternative approach to describe relativistic dynamics,
namely light-front quantum field theory. For extensive discussions
of both its achievements and drawbacks the reader is referred to
one of the reviews on the subject, e.g.\
\cite{burkardt:1996e,brodsky:1997,yamawaki:1998,heinzl:2000}.

Given a Hamiltonian (light-front or otherwise) one can of course
develop perturbation theory without referring to its covariant
version (see e.g.\ Weinberg's text \cite{weinberg:1995}). This
results in (light-front) Hamiltonian perturbation theory.
Light-front Feynman rules were first formulated by Kogut and Soper
for QED \cite{kogut:1970} (see also \cite{brodsky:1973b}). For QCD
they may be found in \cite{lepage:1980}.

Early on it was attempted to show that the covariant perturbation
theory derived from quantisation on null planes coincides with the
one based on canonical quantisation on entirely spacelike
hyperplanes, $t = const$. The idea was to prove that the two Dyson
series for the $S$-matrix coincide no matter if one chooses time
ordering with respect to $t$ or $x^\p$,
\be \label{2S}
  T \exp \left( -i \int dt \, P^0_\mathrm{int} \right) = T_\p \exp
  \left( -i \int dx^\p \, P^\m_{\mathrm{int}} \right) \; .
\ee
Both interaction Hamiltonians, $P^0_\mathrm{int}$ and
$P^\m_{\mathrm{int}}$, are defined in the interaction pictures
associated with the respective choice of time. To actually confirm
the identity (\ref{2S}) is a nontrivial task as $t$ and $x^\p$,
$P^0_\mathrm{int}$ and $P^\m_{\mathrm{int}}$ and the time-ordering
prescriptions are distinct from each other. Obviously, the
differences have to conspire in such a way as to yield overall
cancelation. Using Schwinger's functional formalism this has been
achieved in a series of papers by Yan et al.
\cite{chang:1973a,chang:1973b,yan:1973a,yan:1973b} (see also
\cite{sokolov:1979} and \cite{kvinikhidze:1989} for related
attempts). However, from time to time concerns have been raised as
to whether the proof is water tight. First, it is based on
perturbative reasoning even if formally all orders have been
included. Second, and somewhat related, the notorious singularity
at vanishing longitudinal momentum \cite{maskawa:1976a} and the
ensuing issue of light front `zero modes' has not been taken into
account (see \cite{heinzl:2003} for a recent discussion). Third,
and most important, the questions of regularisation and
renormalisation have only been touched upon.

There is, however, an alternative approach to noncovariant and, in
particular, light-front perturbation theory which guarantees
equivalence with the covariant formulation. In this approach,
pioneered by Chang and Ma \cite{chang:1969b}, one starts with the
covariant Feynman diagrams and performs the integration over the
energy variable $k^\m$ by means of residue techniques.
(Alternatively, one may integrate over the scalar $k^2$ as
suggested in \cite{ritus:1972,schmidt:1974}). Compared to the
instant form, the difference in the number of diagrams now arises
because of the different number of poles in the complex $k^0$ or
$k^\m$ planes, respectively, cf.\ (\ref{P0POLES}) vs.\
(\ref{PMINPOLE}). If the integrations are done properly,
equivalence of covariant and light-front perturbation theory is
guaranteed. However, there are quite a few subtleties involved.
Only recently has it been pointed out that some integrands may not
decay fast enough for large $k^\m$ so that there are `arc
contributions' from the integration contour \cite{bakker:2005b}.
For the time being these represent but the last in a list of
`spurious singularities' that have been thoroughly discussed by
Bakker, Ji and collaborators \cite{bakker:2000a,bakker:2001} (for
an overview of their earlier work, see \cite{bakker:2000b} and
references therein).

Somewhat more relevant to our discussion, however, are
singularities arising whenever the external momenta of a diagram
vanish. A prominent example are the vacuum diagrams which have
already been analysed by Chang and Ma \cite{chang:1969b} and in
more detail by Yan \cite{yan:1973b}. These authors have noted
that, for vacuum diagrams, Feynman integrands typically involve
contributions proportional to $\delta(k^\p)$. Mathematically, the
issue has been summarised in Yan's formula \cite{yan:1973b},
\be \label{YAN}
  \int \frac{dk^\m}{2\pi i} \frac{1}{(k^\p k^\m - M^2 + i
  \epsilon)^2} = \frac{\delta(k^\p)}{M^2} \; .
\ee
This may be shown straightforwardly using Schwinger's
parametrisation to exponentiate denominators. Note that the
integral (\ref{YAN}) cannot be obtained via the standard residue
techniques as the pole in the complex $k^\m$ plane, $k^\m = [k^\m]
\sim 1/k^\p$ is shifted to infinity for $k^\p \to 0$. Hence, the
contour can no longer be closed around the pole implying a delta
function singularity.

Yan's formula has not been appreciated much throughout the
literature although similar delta function `divergences' have been
noted from time to time
\cite{brodsky:1973b,burkardt:1991b,griffin:1992c,burkardt:1993b,bakker:2000b}.
The fact that (\ref{YAN}) seems to be of relevance only for vacuum
diagrams rather than genuine scattering amplitudes or $n$-point
functions apparently justifies this neglect. However, in
\cite{heinzl:2003} and \cite{heinzl:2002} we have pointed out that
a slight generalisation of Yan's formula to arbitrary powers $\nu
\ne 1$ of the denominator,
\be \label{YANNU}
  \int \frac{dk^\m}{2\pi i} \frac{1}{(k^\p k^\m - M^2 + i
  \epsilon)^\nu} = \frac{(-1)^\nu}{\nu - 1} \frac{\delta(k^\p)}{(M^2)^{\nu -
  1}} \; ,
\ee
basically calculates the $\nu$th term in the perturbative
expansion of the one-loop effective potential. From another point
of view (\ref{YANNU}) essentially yields the analytic
regularisation \cite{speer:1968} of the scalar tadpole diagram (in
$d=2$), the divergence being exhibited as the pole at $\nu = 1$.

Admittedly, the effective potential is a physical quantity of
somewhat restricted importance, at least in the context of
scattering theory. In this paper, however, we will argue that the
generalised Yan formula (\ref{YANNU}) is actually of a much
broader relevance than expected hitherto. It turns out that it may
be used to calculate \textit{any} one-loop $n$-point function, the
number of external legs as well as the particle types involved
being arbitrary.

Before going \textit{in medias res} let us conclude this
introduction with a `philosophical' remark. Some people regard
light-front perturbation theory as obtained via energy
integrations (`projection') from covariant diagrams as a `cheat'
or at least as incomplete. This objection is justified to some
extent. The method certainly cannot show that light-front
\textit{Hamiltonian} perturbation theory based on the Poincar{\'e}
generator $P^\m$ as in (\ref{2S}) is equivalent to covariant
perturbation theory -- for the simple reason that it does not
yield the light-front Hamiltonian $P^\m$. Our point of view here
is that the `projected' light-front perturbation theory as
developed in this paper represents a \textit{bona fide} (UV
finite!) approach that should correspond to some (yet unknown)
proper light-front Hamiltonian. The method to be presented might
even contain sufficient information to determine this Hamiltonian
although we will not address this question in this paper. Finding
the `correct' Hamiltonian will presumably require the solution of
both the UV renormalisation and IR zero mode problems which seem
to be entangled with each other (see \cite{heinzl:2003} for a
recent discussion). It should also be pointed out that it is a
well-established and valid procedure to define a quantum field
theory in terms of Feynman rules rather than a Hamiltonian or an
action, in particular if the latter are not known explicitly. A
prominent recent example of this kind are the MHV rules for which
no effective action is known at present (see, however, the
light-front approach in \cite{mansfield:2005,ettle:2006}).

The paper is organised as follows. In Section 2 we apply Yan's
formula to general scalar $n$-point functions. The following two
sections are devoted to the analysis of the special cases of two-
and three-point functions, respectively. We show explicitly how to
recover the standard expressions of light-front perturbation
theory, however, with the bonus of UV finiteness inherited from
the original, dimensionally regulated diagrams. We conclude in
Section 5 with a summarising discussion. Two appendices provide
technical details and deal with ramifications that lead somewhat
off the main line of our manuscript but might still be of
interest.

\section{\label{sec:n-pt} The scalar N-point function}

\begin{figure}[!h]
\includegraphics[scale=0.85]{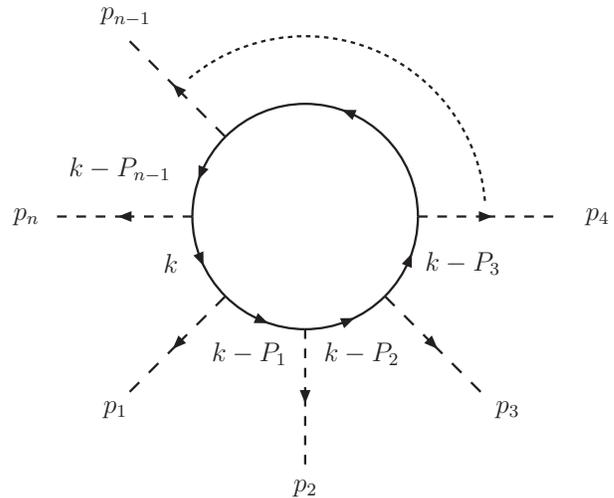}
\caption{\label{fig:n-point} Momentum assignment for the one-loop
$n$-point function.}
\end{figure}

Let us get started with a general discussion of the one-loop
$n$-point function displayed in Fig.~\ref{fig:n-point}. For
reasons of simplicity we have chosen the underlying theory to
involve only scalars with a cubic interaction of $\phi^2 \sigma$
type.

The external momenta $p_1, \ldots , p_n$ (corresponding to the
$\sigma$-field) are all flowing out of the diagram and hence sum
up to zero. All internal lines are assumed to correspond to scalar
propagators with equal masses, the $ith$ one being
\be \label{prop}
  G_i (k) \equiv \frac{1}{(k - P_i)^2 - m^2} \; ,
\ee
where we have suppressed the (causal) pole prescription. The loop
momentum to be integrated is $k$ and
\be
  P_0 \equiv 0 \; , \quad P_i \equiv p_1 + \ldots + p_i \; , \quad
  i = 1 , \ldots , n-1 \; .
\ee
In terms of the propagators (\ref{prop}) the graph of
Fig.~\ref{fig:n-point} represents the momentum integral
\be \label{IND1}
  I_{n,d} \equiv \int \frac{d^d k}{(2\pi)^d} \prod_{i=0}^{n-1}
  \frac{1}{(k - P_i)^2 - m^2} \; .
\ee
The denominators appearing in the integrand may be combined by
introducing the usual Feynman parameters $x_0, \ldots \ x_{n-1}$
(with $x_0 = 1 - \sum x_i$, see App.~\ref{APP:FP}) leading to the
compact expression
\be \label{IND2}
  I_{n,d} =  \int_0^1  [d^{n-1} x]  \int
  \frac{d^d \ell_n}{(2\pi)^d} \frac{\Gamma(n)}{(\ell_n^2 -
  \Delta_n)^n} \; ,
\ee
which will henceforth be referred to as the Feynman
representation. In (\ref{IND2}) we have defined a measure on the
$(n-1)$-simplex,
\be
  [d^{n-1} x] \equiv d^{n}x \; \delta(1 - x_0 - \sum_i x_i) \; ,
\ee
a shifted 4-momentum
\be \label{ELLN}
  \ell_n (\vc{x}) \equiv k - \sum_i x_i P_i \; ,
\ee
and an effective mass term
\be \label{DELTAN}
  \Delta_n (\vc{x}) \equiv m^2 - \sum_i x_i \bar{x}_i P_i^2 + 2
  \sum_{j<i} x_i x_j P_i \cdot P_j  \; ,
\ee
employing the abbreviations $\vc{x} \equiv (x_1 , \ldots
x_{n-1})$, $d^{n}x \equiv dx_0 \ldots dx_{n-1}$, $\bar{x}_i \equiv
1 - x_i$, and all unconstrained summations extending from $i=1$ to
$n-1$. Note that $\Delta_n$ is a quadratic form in the $x_i$ which
may be written in condensed notation \cite{ferroglia:2002},
\be
  \Delta_n (\vc{x}) = (\vc{x}, H \vc{x}) + 2 (K, \vc{x}) + L \; ,
\ee
with the coefficients encoded in the Lorentz invariants
\be
  H_{ij} \equiv - P_i \cdot P_j \; , \quad K_{i} \equiv P_i^2 \; ,
  \quad L \equiv m^2 \; .
\ee
The momentum integral in (\ref{IND2}) can be found in standard
texts (see e.g.\ \cite{peskin:1995}),
\be \label{INTFORM}
  \int d^d \ell_n \, \frac{1}{(\ell_n^2 - \Delta_n)^n} = i (-1)^n
  \frac{\Gamma(n - d/2)}{\Gamma(n)} \frac{\pi^{d/2}}{\Delta_n^{n - d/2}}
  \; ,
\ee
whereupon (\ref{IND2}) becomes
\be \label{IND3}
  I_{n,d} = i \frac{(-1)^n}{(4\pi)^{d/2}} \Gamma(n - d/2)
  \int_0^1 \frac{[d^{n-1} x]}{\Delta_n^{n - d/2}(\vc{x})} \; .
\ee
The hard part still to be done is to integrate over the $n-1$
independent Feynman parameters with the integrand being a
complicated rational function of the $x_i$. There is no general
formula available and, for standard applications, not really
required as one is usually only interested in the behaviour of the
integrals for small deviation $\epsilon$ from four dimensions, $d
= 4 - 2\epsilon$. In App.~\ref{APP:DIRICHLET} we point out the
possibility to evaluate the parameter integrals as statistical
averages.

Our next goal is to interpret the $x_i$ in terms of light-front
momentum fractions in the spirit of Weinberg \cite{weinberg:1966a}
rather than to perform the Feynman parameter integrals. The
crucial observation is that the momentum integral (\ref{IND2}) may
be evaluated in terms of light-front coordinates making use of the
generalised Yan formula (\ref{YANNU}). To this end we rewrite
(\ref{IND2}) using $d^d \ell = (1/2)\, d^{d-2}\ell_\perp \,
d\ell^\p d\ell^\m$,
\be \label{INDLC1}
  I_{n,d} = \int\limits_0^1 [d^{n-1} x] \int \dlp
  \frac{d\ell^\p d\ell^\m}{8\pi^2}
  \frac{\Gamma(n)}{(\ell^\p \ell^\m - M_n^2)^n} \; ,
\ee
where we have omitted the subscripts $n$ of the integration
variables and introduced the effective `transverse mass'
\be \label{MN}
  M_n^2 (\ell_\perp, x) \equiv \ell_{n,\perp}^2 + \Delta_n (\vc{x}) \;
  .
\ee
The $\ell^\m$-integral is now directly amenable to the generalised
Yan formula, the subsequent $\ell^\p$-integration over
$\delta(\ell^\p)$ being trivial. We therefore end up with the
integral representation
\be \label{WEINBERGREP}
  I_{n,d} = i \frac{(-1)^n}{4\pi}  \int\limits_0^1 [d^{n-1}x] \, \int
  \frac{d^{d-2}\ell_{\perp}}{(2\pi)^{d-2}}
  \frac{\Gamma(n-1)}{(M_n^2)^{n-1}} \; .
\ee
The following remarks are in order. The final integral
(\ref{WEINBERGREP}) consists of $n-1$ integrations over Feynman
parameters which are the same as in the covariant expression
(\ref{IND2}) and a \textit{Euclidean} integral over transverse
momenta in dimension $d-2$. This is the structure already
discovered by Weinberg for $d=4$ and $n=2$ \cite{weinberg:1966a}.
It therefore seems appropriate to refer to the integral
(\ref{WEINBERGREP}) as being in the `Weinberg representation'. Its
most important property presumably is its manifest Lorentz
invariance as the sole dependence of $\Delta_n$ on external
momenta is in terms of the invariants $P_i^2$ and $P_i \cdot P_j$,
cf.\ (\ref{DELTAN}).

Note furthermore that, throughout the derivation leading from
(\ref{IND2}) to (\ref{WEINBERGREP}), every expression was finite
by virtue of dimensional regularisation (dimReg). This is true in
particular for the final expression (\ref{WEINBERGREP}) where the
original regularisation has resulted in \textit{transverse}
dimReg, originally suggested by Casher \cite{casher:1976}. (In
this paper we do not discuss infrared singularities which, for
$m=0$, show up as endpoint singularities as $x \to 0$ or 1.)
Finally, it is easy to see using (\ref{INTFORM}) that performing
the transverse momentum integration in (\ref{WEINBERGREP}) exactly
reproduces (\ref{IND3}).

The next task is to relate the Weinberg representation
(\ref{WEINBERGREP}) to the `light-front representation', i.e.\ the
standard integrals encountered in light-front perturbation theory
with energy denominators as given in (\ref{PMINPOLE}). This is the
topic of the next sections.

\section{\label{sec:2-pt}2-point function}

\begin{figure}[!h]
\includegraphics[scale=0.85]{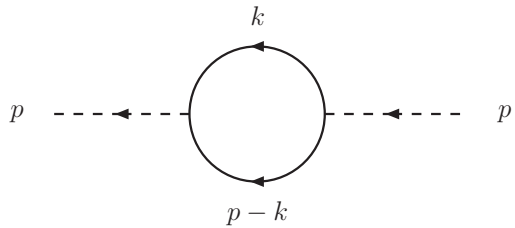}
\caption{\label{fig:2-point} Momentum assignment for the one-loop
$2$-point function.}
\end{figure}

It seems wise to begin with the simplest case, the scalar 2-point
function, i.e.\ $n=2$ (see Fig.~\ref{fig:2-point}). In agreement
with standard conventions we set $p_1 = - p_2 \equiv p$, $x_1
\equiv x = 1 - x_2$ and obtain for (\ref{ELLN}) and
(\ref{DELTAN}),
\bea
  \ell_2 &=& k - x p \; , \label{ELL2} \\
  \Delta_2 &=& m^2 - x \bar{x} \, p^2 \label{DELTA2} \; .
\eea

We thus consider the integral
\be \label{I2COV}
 I_{2,d} =  \int_0^1 dx \, \int \frac{d^d
  \ell_2}{(2\pi)^d} \frac{1}{(\ell_2^2 - \Delta_2)^2}
\ee
and in particular its Weinberg representation,
\be \label{I2LF}
  I_{2,d} = \frac{i}{4\pi}  \int_0^1 dx \, \int
  \frac{d^{d-2}\ell_{\perp}}{(2\pi)^{d-2}}
  \frac{1}{M_2^2} \; ,
\ee
with $M_2^2$ explicitly given by
\be \label{M2EXPL}
  M_2^2 (x,p) = (k_\perp - x p_\perp)^2 + m^2 - x \bar{x} p^2 \; ,
\ee
according to (\ref{MN}), (\ref{ELL2}) and (\ref{DELTA2}).

\begin{figure}[!h]
\includegraphics[scale=0.85]{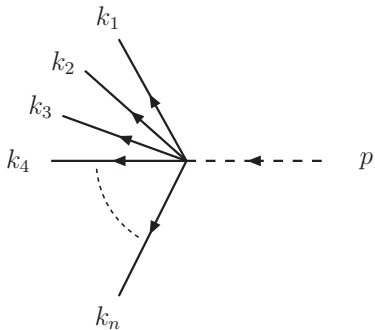}
\caption{\label{fig:LFamp} $n$-particle light-front amplitude
corresponding to an $n$th order scalar vertex.}
\end{figure}

Yan's formula (\ref{YAN}) tells us that the
$\ell_2^\p$-integration leading from (\ref{I2COV}) to (\ref{I2LF})
is localised at $\ell_2^\p = 0$ which according to (\ref{ELL2})
implies
\be \label{XDX}
  x = k^\p / p^\p \; , \quad dx = dk^\p / p^\p \; ,
\ee
identifying $x$ straightforwardly as a longitudinal momentum
fraction. We can thus trade the single $x$-integration in
(\ref{I2LF}) for a $k^\p$-integration ranging from 0 to $p^\p$. To
exhibit the light-front energy denominators we have derived the
algebraic identity
\be \label{M2}
  M_2^2 = - \frac{k^\p (p^\p - k^\p)}{p^\p} \Big\{ p^\m - [k^\m] -
  [p^\m - k^\m] \Big\} \; ,
\ee
which is proven by simply working out the right-hand side and
comparing with (\ref{M2EXPL}). It turns out to be useful to
abbreviate light-front denominators by means of a bra-ket
notation,
\be \label{BRACKET}
  \bra k_1, \ldots , k_n \, | \, p \ket \equiv
  \frac{(p^\p)^{n-1}}{k_1^\p \ldots k_n^\p \, \Big\{ p^\m - [k_1^\m]
  - \ldots - [k_n^\m] \Big\}} \; .
\ee
Here $p$ denotes the total momentum which is distributed among the
$n$ intermediate states labelled by their on-shell momenta $k_1,
\ldots , k_n$. The quantity (\ref{BRACKET}) has dimensions of
inverse mass squared. It may be interpreted as the perturbative
light-front amplitude for a particle with incoming (off-shell)
momentum $p$ to consist of $n$ constituents of momenta $k_i$ (see
Fig.~\ref{fig:LFamp}) and hence represents an off-shell extension
of an $n$-particle light-front wave function.

For scalar fields in $d > 2$ the number $n$ will in general not
exceed 4 as there are no high-order renormalisable vertices. The
situation, however, is different for super-renormalisable theories
in $d=2$, e.g.\ for the sine-Gordon model
\cite{griffin:1992c,burkardt:1993b}.

Using (\ref{BRACKET}) the propagator (\ref{PMINPOLE}) can be
written as a `single-particle' amplitude,
\be \label{PROPREP}
  \frac{1}{p^2 - m^2 + i \epsilon} = \frac{1}{p^\p} \frac{1}{p^\m -
  [p^\m]} \equiv \bra p \, | \, p \ket \; ,
\ee
and the two-particle amplitude is the inverse of the transverse
mass (\ref{M2})
\be \label{M2SHORT}
   \bra k , p-k \, | \, p \ket \equiv -1/M_2^2 \; .
\ee
Plugging (\ref{M2SHORT}) and (\ref{XDX}) into (\ref{I2LF}) finally
yields the desired light-front representation,
\be \label{I2LFDEN}
  I_{2,d} = -i \int_0^{p^\p} \frac{dk^\p}{4\pi p^\p} \int
  \frac{d^{d-2}k_{\perp}}{(2\pi)^{d-2}} \bra k , p-k \, | \, p
  \ket \; .
\ee
It is worth reemphasising that this is strictly identical to the
covariant expression (\ref{I2COV}) and the Weinberg representation
(\ref{I2LF}). Using the general result (\ref{IND3}) the 2-point
function becomes
\be \label{I2EVAL}
  I_{2,d} = i \, \frac{\Gamma(2-d/2)}{(4\pi)^{d/2}} \int_0^1
  \frac{dx}{(m^2 - x \bar{x} \, p^2)^{2-d/2}} \; ,
\ee
which makes its Lorentz invariance manifest. Weinberg in
\cite{weinberg:1966a} has basically obtained (\ref{I2LF}) in $d=4$
and subsequently shown that the derivatives of (\ref{I2LF}) and
(\ref{I2EVAL}) with respect to $p^2$ (which are finite in $d=4$)
coincide. He did \textit{not} relate these integrals to the
standard light-front representation (\ref{I2LFDEN}) which was
still awaiting its discovery at that time \cite{chang:1969b}.

The approach presented in this section suggests an interesting
route to evaluating one-loop diagrams in light-front perturbation
theory. If one manages to rewrite any light-front representation
with all its energy denominators as a Weinberg representation
using transverse dimReg one has achieved an elegant way of both
doing the transverse integrations and proving Lorentz invariance.
All UV divergences should be regularised and one has only to deal
with the same IR divergences as are present in the covariant
diagram.

\section{\label{sec:3-pt}3-point function}

According to our general formula (\ref{IND2}) the scalar 3-point
function has the covariant Feynman representation
\be \label{}
  I_{3,d} =  \int_0^1 dz \int_0^{1-z} dz' \, \int \frac{d^d
  \ell_3}{(2\pi)^d} \frac{2}{(\ell_3^2 - \Delta_3)^3} \; .
\ee
The Feynman parameters have been renamed as $z$ and $z'$ while
$\ell_3$ and $\Delta_3$ follow from (\ref{ELLN}) and
(\ref{DELTAN}). The latter will be stated explicitly below once we
have chosen particular kinematics. Yan's formula implies the
Weinberg representation
\be \label{I3WBG}
  I_{3,d} = - \frac{i}{4\pi} \int\limits_0^1 [d^2 z] \int \dlp
  \frac{1}{[M_3^2 (\vc{z}, \ell_\perp) ]^2} \; ,
\ee
with $M_3^2$ as defined in (\ref{MN}). It is not entirely
straightforward to transform (\ref{I3WBG}) into its the
light-front representation. One would have to decompose $M_3^2$
into a sum of energy denominators describing the intermediate
2-particle states. We found it actually simpler to reverse the
order of Section \ref{sec:2-pt} and work `backwards' from the
light-front to the Weinberg representation. Still, for the most
general 3-point function the associated light-front representation
becomes quite tedious to determine. To see what is involved let us
rewrite the covariant Feynman diagram using the denominator
replacement (\ref{PROPREP}) which yields
\be \label{I3}
  I_{3,d} = \int \dkp \int
  \frac{dk^\p dk^\m}{8\pi^2} \prod_{i=0}^2
  \bra k+P_i \, | \,  k+P_i \ket \; .
\ee
In order to perform the $k^\m$-integration using residue
techniques one has to determine the location of the poles in the
complex $k^\m$-plane. Reinstating the $i\epsilon$ prescriptions
the sign of the poles' imaginary part is given by $\sgn(k^\p +
P_i^\p)$. One thus has to consider all orderings of the
longitudinal momenta $-P_i^\p$ and check whether closing a contour
yields a contribution \cite{bakker:2000b}. This is the case if the
real $k^\m$-axis is pinched between poles. Closer examination
reveals that pinching can only happen  if the following `pinching
condition' is satisfied,
\be \label{PINCHCOND}
  \min_i (-P_i^\p) < k^\p < \max_i (-P_i^\p) \; .
\ee
Otherwise, all poles will be located either above or below the
real axis and the $k^\m$-integration yields zero. We thus conclude
that the integration region for $k^\p$ will be finite with the
boundaries given by (\ref{PINCHCOND}). This confirms and
generalises the original findings of \cite{chang:1969b}. Shifting
the integration variable $k^\p$ appropriately one can always
achieve a standard integration range for longitudinal momentum
from zero up to some maximum value (see examples below).

For the 3-point functions one \textit{a priori} has $3! = 6$
$P_i^\p$-orderings. Choose one of these and let $k^\p$ gradually
increase from negative towards positive infinity. To the left of
the region (\ref{PINCHCOND}) one gets zero as all poles are on the
same side of the real axis. Pole after pole will cross the axis
whenever $k^\p = -P_i^\p$. The final crossing will again result in
all poles being to one side and hence a vanishing contribution.
For $n=3$ we start with the three poles, say, above the real axis
and none below, a configuration which we denote by $(3|0)$.
Increasing $k^\p$ we successively obtain configurations $(2|1)$,
$(1|2)$ and $(0|3)$. Only those with nonzero entries contribute,
i.e.\ $(2|1)$ and $(1|2)$. Thus, for a general $n$-point we expect
$n-1$ non-vanishing pole configurations such that the total number
of light-front integrals contributing should be
\be \label{NN}
  N_n = (n-1) \, n! \; .
\ee
Obviously, the number of integrals grows somewhat stronger than
factorially. For the two-point function of the previous section,
there are \textit{a priori} two contributions stemming from the
`pinching intervals' $0 < k^\p < p^\p$ and $p^\p < k^\p < 0$.
However, both give rise to the same light-front integral
representation, as both regions imply $0 < x < 1$. That's why in
the end there is only one integral left.

For $n=3$ the counting rule (\ref{NN}) implies a total of 12
contributions. Again, depending on the symmetries and kinematics
involved, some of those may still vanish. To simplify things we
will exploit this fact by considering the popular choice of
\textit{form factor} kinematics. This has been studied extensively
in recent years leading to a vast amount of literature. It is
impossible to give a complete account of the latter and we only
refer the reader to a representative list of references
\cite{brodsky:1973b,bakker:2001,bakker:2005b,frankfurt:1979,frederico:1992,melikhov:2001}
where the relation between the covariant triangle diagram and
light-front perturbation theory has been investigated.

\begin{figure}[!h]
\includegraphics[scale=0.8]{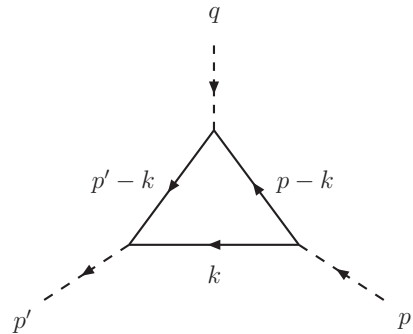}
\caption{\label{fig:3-point} Momentum assignment for the one-loop
$3$-point function (form factor kinematics).}
\end{figure}

Form factor kinematics amounts to setting
\be \label{REPLACE3}
  p_1 = -p \; , \quad p_2 = p' \; , \quad p_3 = -q \; ,
\ee
and replacing $k \to k - p$. $q$ is interpreted as  the probe
momentum transfer (Fig.~\ref{fig:3-point}). Strictly speaking, for
a form factor $p$ and $p'$ are on-shell which we do not assume for
the time being.

The $n=3$ abbreviations then read explicitly
\bea
  M_3^2 &=& \ell_{3,\perp}^2 + \Delta_3 \; , \\
  \ell_3 &=& k - z p - z' p' \; , \label{ELL3} \\
  \Delta_3 &=& m^2 - z \bar{z} p^2 - z' \bar{z}' p^{\prime 2} +
  2zz' p \cdot p' \; .
\eea
In what follows we will further assume that the momentum transfer
$q$ has $q^\p \ge 0$ which includes the simplest case, namely the
Drell-Yan-West frame, $q^\p = 0$. This implies the longitudinal
momentum ordering
\be
  0 < p^\p \le p^{\prime \p} = p^\p + q^\p \; ,
\ee
which in turn determines the location of the $k^\m$-poles.
Choosing this particular ordering reduces the number of
light-front integrals from 12 to 2, which, after residue
integration in (\ref{I3}) become
\bea \label{I3LF}
  I_{3,d} &=& \frac{-i}{4\pi} \int \dkp \int\limits_0^{p^\p}
  \frac{dk^\p k^\p}{p^\p p^{\prime \p}} \bra p' \, | \, k , p'-k
  \ket \bra k, p-k \, | \, p \ket \nn \\
  &-& \frac{i}{4\pi} \int \dkp \int\limits_0^{q^\p} \frac{dk^\p
  k^\p}{q^\p p^{\prime \p}}
  \bra p' \, | \, k , p'-k \ket \bra k, q-k \, | \, q \ket  \nn \\
  &\equiv& I_{3,d}^{(\mathrm{v})} + I_{3,d}^{(\mathrm{nv})} \; .
\eea

\begin{figure*}
\includegraphics[scale=0.85]{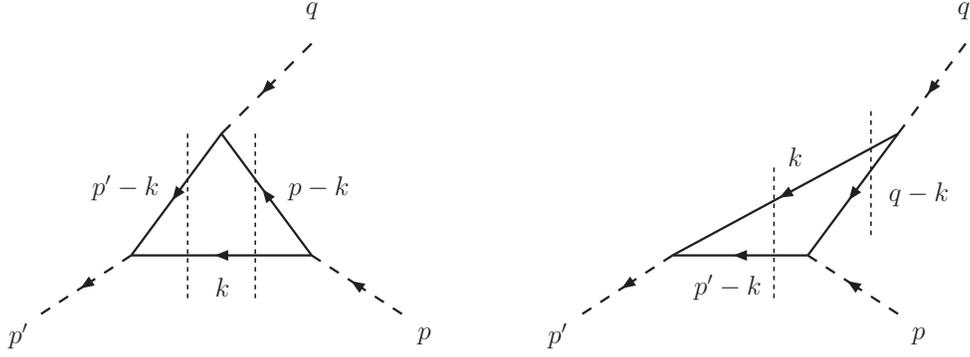}
\caption{\label{fig:3-pointLF} The two contributions to the
one-loop $3$-point function (form factor kinematics) in
light-front perturbation theory. Dashed vertical lines indicate
two-particle denominators. Left: valence contribution; right:
nonvalence contribution}
\end{figure*}

Note the $k^\p$-measures which will become important in a moment.
The second integral in (\ref{I3LF}) is obtained from the first
upon replacing $p \leftrightarrow q$ (which leaves $p' = p+q$
invariant). In Fig.~\ref{fig:3-pointLF} we have depicted the
associated diagrams usually referred to as the `valence' and
`nonvalence contribution', hence the superscripts in (\ref{I3LF}).
The light-front time direction is pointing from right to left.
Hence, in the second diagram the first vertex (labelled by $q$)
corresponds to pair creation so that the $p$-vertex corresponding
to the incoming particle cannot be interpreted as a light-front
wave function with $p$ being on shell \cite{sawicki:1992}.

The question now is how to obtain the Weinberg representation
(\ref{I3WBG}) from the light-front one given by (\ref{I3LF}). The
answer is technically somewhat tricky. One first generalises
(\ref{XDX}) to $n=3$ by defining the longitudinal momentum
fractions
\be
  x \equiv k^\p / p^\p \; , \quad x' \equiv k^\p / p^{\prime \p}
  \; .
\ee
Naturally one is tempted to follow the case $n=2$ and identify $x$
and $x'$ directly with $z$ and $z'$ from (\ref{I3WBG}).
Unfortunately, this does not make sense as the former are not
independent,
\be \label{XXP}
  x' = \frac{p^\p}{p^{\prime\p}} x \equiv \kappa x \; .
\ee
However, we may follow the treatment of the 2-point function in
rewriting the light-front denominators in (\ref{I3LF}) by means of
(\ref{M2SHORT}). This yields the fairly compact expression
\be \label{I3MM}
  I_{3,d}^{\mathrm{(v)}} = - \kappa \frac{i}{4\pi} \int \dkp
  \int\limits_0^1 \frac{dx \, x}{M_2^2 M_2^{\prime 2}} \; ,
\ee
with $M_2^2$ as given in (\ref{M2EXPL}) and $M_2^{\prime 2} \equiv
M_2^2 (x' , p')$. Note the appearance of the factor $\kappa = p^\p
/ p^{\prime \p}$ in front of the integral. We still have only one
Feynman parameter $x$ in (\ref{I3MM}), but this is easily remedied
by combining the two denominators with a second parameter $y$,
\be \label{I32FP}
  I_{3,d}^{\mathrm{(v)}} = -  \frac{i\kappa}{4\pi} \int \dkp
  \int\limits_0^1 dx x \int\limits_0^1 dy \frac{1}{\Big[ y
  M_2^2 + \bar{y} M_2^{\prime 2}\Big]^2} \; .
\ee
This starts to look promising and indeed the `miracle' happens. A
lengthy but straightforward calculation shows the identity
\be \label{M23}
  y \, M_2^2 (x,p) + \bar{y} M_2^2 (x' , p') = M_3^2 (z,z',
  \ell_{3,\perp}) \; ,
\ee
if the following identifications are made,
\be \label{XYZ}
  z \equiv x y \; , \quad z' \equiv x \bar{y} \;
\ee
and $\ell_3$ from (\ref{ELL3}) is used. Working out the Jacobian
yields the additional relation
\be
  dz \, dz' =  x \, dx \, dy  \; .
\ee
This is exactly the measure appearing in expression (\ref{I32FP})
which thus turns into the Weinberg representation
\be \label{I3V}
  I_{3,d}^{\mathrm{(v)}} = -  \frac{i\kappa}{4\pi} \int\limits_0^1
  dz \int\limits_0^{1-z} dz' \int \dlp \frac{1}{[M_3^2 (z,z',
  \ell_\perp)]^2} \; .
\ee
For the nonvalence contribution of (\ref{I3LF}) we simply replace
$p$ by $q$ wherever the former appears. This leads to the same
integrals as in (\ref{I3V}) the only difference being the
prefactor replacement $\kappa \to 1 - \kappa$. Adding both
contributions we finally end up with
\be
  I_{3,d} = -  \frac{i}{4\pi} \int\limits_0^1 dz \int\limits_0^{1-z}
  dz' \int \dlp \frac{1}{[M_3^2 (z,z', \ell_\perp)]^2} \; .
\ee
Note that $\kappa$ is invariant under longitudinal boosts under
which $p^\p \to \lambda p^\p$. Hence, both contributions,
$I_{3,d}^{\mathrm{(v)}}$ and $I_{3,d}^{\mathrm{(nv)}}$ are
separately boost invariant. Complete Lorentz invariance, however,
is only achieved by adding both terms so that the $\kappa$
dependence cancels.

\section{Discussion and conclusion}

Let us summarise our findings. We have represented the Feynman
diagram $I_{n,d}$ for a general scalar $n$-point function in
basically three different ways all of which are \textit{strictly
finite} by means of dimReg. The different representations are:

\noindent \textit{1. Feynman representation}
\be \label{FREP}
  I_{n,d} =  \int_0^1  [d^{n-1} x]  \int
  \frac{d^d \ell}{(2\pi)^d} \frac{\Gamma(n)}{(\ell^2 -
  \Delta_n)^n} \; .
\ee
This is the standard textbook representation obtained after
combining denominators with Feynman parameters. The momentum
integral is straightforwardly done with formula (\ref{INTFORM}).

\noindent \textit{2. Weinberg representation}
\be \label{WREP}
  I_{n,d} = i \frac{(-1)^n}{4\pi}  \int\limits_0^1 [d^{n-1}x] \, \int
  \frac{d^{d-2}\ell_{\perp}}{(2\pi)^{d-2}}
  \frac{\Gamma(n-1)}{(\ell_{\perp}^2 + \Delta_n)^{n-1}} \; .
\ee
This representation is obtained from (\ref{FREP}) by performing
the $\ell^\m$-integration using Yan's formula (\ref{YANNU}).
Afterwards the integral is localised at vanishing longitudinal
momentum, $\ell^\p = 0$, by means of a delta function which is
trivially integrated in turn. Note that (\ref{WREP}) is manifestly
invariant as $\Delta_n$ only depends on Lorentz scalars. Again,
the momentum integration may be done with (\ref{INTFORM}). A
precursor of (\ref{WREP}) has already been obtained by Weinberg
\cite{weinberg:1966a}.

\noindent \textit{3. Light-front representation}

For the two- and three-point functions ($n=2$ and 3, respectively)
we have succeeded in explicitly relating the Weinberg
representation (\ref{WREP}) to its light-front analogue. That this
can be done, in particular for the nontrivial case $n=3$, is one
of the main results of this paper. Explicitly, one has for $n=2$,
\be \label{LFREP2}
  I_{2,d} = -i \int_0^{p^\p} \frac{dk^\p}{4\pi p^\p} \int
  \frac{d^{d-2}k_{\perp}}{(2\pi)^{d-2}} \bra k , p-k \, | \, p
  \ket \; .
\ee
and for $n=3$,
\bea
  I_{3,d} &=& \frac{-i}{4\pi} \int \dkp \int\limits_0^{p^\p}
  \frac{dk^\p k^\p}{p^\p p^{\prime \p}} \bra p' \, | \, k , p'-k
  \ket \bra k, p-k \, | \, p \ket \nn \\
  &+& (p \leftrightarrow q) \; .
\eea
The technical challenge is to transform Feynman parameters into
longitudinal momentum fractions (which is straightforward only for
$n=2$) and to factorise the covariant denominator in (\ref{WREP})
in terms of light-front energy denominators, abbreviated via the
bracket notation $\bra k_1, \ldots k_n | p \ket$ from
(\ref{BRACKET}).

Obviously, one would like to have a general result for the
light-front representation of $I_{n,d}$ for arbitrary $n$. While
this might be rather involved, it seems reasonable to expect at
least a recursion type formula relating the denominator
expressions $M_n^2$ to $M_{n-1}^2$, cf.\ (\ref{M23}), as well as
relations between integration parameters analogous to (\ref{XYZ}).
Clearly, this deserves further investigation.

As a by-product of this investigation we obtain alternative ways
of regularising perturbative light-front wave functions. The
latter can be read off by writing the form factor integrand of
$I_{3,d}$ (for zero momentum transfer) as a wave function squared
and putting external momenta on-shell, $p^2 = p^{\prime 2} = M^2$.
This yields wave functions \cite{terentev:1976,frankfurt:1979}
\be \label{LCWF}
  \psi (x, \vc{\ell}_\perp) = \frac{N}{M^2 - (\vc{\ell}_\perp^2 +
  m^2)/x\bar{x}} \; ,
\ee
where $N$ is a `normalisation' constant. The use of inverted
commas serves to remind us of the fact that normalisation requires
regularisation, and dimReg indeed does the job for us. At zero
transverse separation (which defines the distribution amplitude),
for instance, we find
\be
  \int d^2 \ell_\perp \, \psi(x, \vc{\ell}_\perp) \sim x \bar{x}
  \, \frac{1}{\epsilon} + \mbox{finite terms} \; ,
\ee
with $\epsilon = 2- d/2$ as usual. The same result is obtained
within analytic regularisation, where the denominator in
(\ref{LCWF}) is raised to power $\nu$ and $\epsilon = \nu - 1$.

With hindsight it is the Weinberg representation (\ref{WREP})
which looks most appealing from a `noncovariant' point of view.
Thus, the intriguing question arises as to whether it is possible
to derive this representation directly from a modified light-front
Lagrangian or Hamiltonian. At present, we do not have a
satisfactory answer but it definitely seems worthwhile to keep
looking for it.

\begin{acknowledgments}
I am grateful to Arsen Khvedelidze, Kurt Langfeld, Martin Lavelle,
and David McMullan for useful discussions during the course of
this work. Special thanks go to Volodya Karmanov for explaining
the difficulties of light-front form factor calculations, to Ben
Bakker for providing me with copies of his work, and to Anton
Ilderton for a critical reading of the manuscript.
\end{acknowledgments}

\appendix

\section{Feynman parameters}
\label{APP:FP}

There are several equivalent ways of representing products of
denominators in terms of Feynman parameters. (The reference to
Feynman is actually misleading, as it was Schwinger who originally
suggested the method. This was explicitly stated by Feynman
himself in \cite{feynman:1949b} and was recently emphasized in
\cite{milton:2006}.)

The most straightforward formula presumably is the following,
\be
  \frac{1}{a_0 a_1 \ldots a_{n-1}} = \int d^n x \, \frac{\Gamma (n)
  \, \delta(1 - \sum_{i=0}^{n-1} x_i)}{ (a_0 x_0 +  \ldots +
  a_{n-1} x_{n-1})^n} \; ,
\ee
with the (flat) measure $d^n x \equiv dx_0 dx_1 \ldots dx_{n-1}$.
The delta function entails that the integration actually extends
over the $(n-1)$-simplex with measure $[d^{n-1}x]$. Depending on
the variables (Feynman parameters) chosen, this measure takes on
many different forms, each with its own integration boundaries.
While proving the validity of the different representations by
induction is straightforward, it is a nontrivial task to find the
variable transformations relating them.

The following representation (see e.g.\ \cite{ho-kim:1998}) was
particularly useful for our purposes,
\begin{widetext}
\be
  \frac{1}{a_0 a_1 \ldots a_{n-1}} = \int\limits_0^1 dx_1
  \int\limits_0^{1-x_1} dx_2 \ldots \int\limits_0^{1-x_1 - \ldots
  -x_{n-2}} dx_{n-1} \frac{\Gamma(n)}{\Big[ a_0 (1-x_1 - \ldots
  -x_{n-2}) + a_1 x_1 + \ldots + a_{n-1} x_{n-1} \Big]^n} \; .
\ee
A different but equivalent way of writing this is (see e.g.\
\cite{jauch:1955}, App.~A5 or \cite{quigg:1983}, App.~B.3)
\be
  \frac{1}{a_0 a_1 \ldots a_{n-1}} = \int\limits_0^1 du_1
  u_1^{n-2} \int\limits_0^{1} du_2 u_2^{n-3} \ldots
  \int\limits_0^{1} du_{n-1} \frac{\Gamma(n)}{\Big[ (a_0 - a_1)
  u_1 \ldots u_{n-1} + (a_1 - a_2) u_1 \ldots u_{n-2} + \ldots +
  a_{n-1} \Big]^n} \; .
\ee
\end{widetext}

\section{Tensor integrals and the Dirichlet distribution}
\label{APP:DIRICHLET}

Within the Passarino-Veltman scheme there is a standard procedure
to reduce tensor integrals (which typically appear for nonzero
spin) to scalar integrals, see e.g.\ \cite{weinzierl:2003}. The
main differences to the integrals encountered so far in this paper
are a shift in dimension, $d \to d'$, and nontrivial exponents in
the denominators,
\be \label{INUI}
  (k - P_i)^2 - m^2 \to [(k - P_i)^2 - m^2]^{\nu_i} \; .
\ee
Effectively, this changes the scalar integrals (\ref{IND1})
according to
\be
  I_{n,d} \to I_{n',d', \ssvc{\nu}} \; ,
\ee
where $\vc{\nu}$ is the vector formed from the exponents $\nu_i$
in (\ref{INUI}). Explicitly, we have instead of (\ref{IND1})
\be \label{INDNU0}
  I_{n,d, \ssvc{\nu}} = \int \frac{d^d k}{(2\pi)^d}
  \prod_{i=0}^{n-1} \frac{1}{[(k - P_i)^2 - m^2]^{\nu_i}} \; ,
\ee
where we omitted the primes on $n$ and $d$ for simplicity. Again,
upon keeping $d$ fixed and setting $\nu_i = 1 + \epsilon$ this
integral may alternatively be viewed as the analytic
regularisation of $I_{n,d}$. We will not go down this road,
however, but rather stick with the dimReg interpretation.

Introducing Feynman parameters as before (\ref{INDNU0}) becomes
\be \label{INDNU1}
  I_{n,d, \ssvc{\nu}} =  \int\limits_0^1 \frac{[d^{n-1}x]}{B(\vc{\nu})} \prod_{i=0}^{n-1}
  x_i^{\nu_i - 1} \int \frac{d^d \ell_n}{(2\pi)^d} \frac{1}{(\ell_n^2 -
  \Delta_n)^{|\nu|}} \; ,
\ee
where $|\nu| \equiv \nu_0 + \nu_1 + \ldots + \nu_{n-1}$ and $B$
denotes the multinomial Beta function,
\be
  B(\vc{\nu}) \equiv \frac{\Gamma(\nu_0) \ldots
  \Gamma(\nu_{n-1})}{\Gamma(|\nu|)} \; .
\ee
Somewhat surprisingly, the integrals (\ref{INDNU1}) have a nice
statistical interpretation. If one introduces the Dirichlet
distribution \cite{devroye:1986,ohagan:2004} on the
$(n-1)$-simplex, which corresponds to a probability density
\be
  \rho_\mathrm{Dir} (\vc{x}, \vc{\nu}) \equiv
  \frac{1}{B(\vc{\nu})} \prod_{i=0}^{n-1} x_i^{\nu_i - 1} \,
  \delta\left( 1 - \sum_{i=0}^{n-1} x_i \right) \; ,
\ee
one may define the expectation values
\be
  \left\langle f(\vc{x}) \right\rangle_{\, \mathrm{Dir}(\ssvc{\nu})}
  \equiv \int\limits_0^1 dx_0 \ldots dx_{n-1} \, \rho_\mathrm{Dir}
  (\vc{x}, \vc{\nu}) \, f(\vc{x}) \; .
\ee
Hence, the integrals (\ref{INDNU1}) are nothing but the
expectation values
\be \label{INDNU2}
  I_{n,d, \ssvc{\nu}} =
  \left\langle \int \frac{d^d \ell_n}{(2\pi)^d} \frac{1}{[\ell_n^2 (\vc{x}) -
  \Delta_n (\vc{x})]^{|\nu|}}
  \right\rangle_{\mathrm{Dir}(\ssvc{\nu})} \; ,
\ee
with the momentum integral to be evaluated according to
(\ref{INTFORM}). We conclude these remarks with the the scalar
integral discussed in the main part,
\be \label{INDUNIFORM}
  I_{n,d} = \left\langle \int \frac{d^d
  \ell_n}{(2\pi)^d} \frac{1}{[\ell_n^2 (\vc{x}) - \Delta_n
  (\vc{x})]^n} \right\rangle_{\mathrm{Dir}(\ssvc{1})} \; .
\ee
and note that Dir$(\vc{1})$ represents the uniform distribution on
the simplex.

Obviously, for a reasonably large number of external legs it
should be feasible to evaluate the integrals (\ref{INDNU2}) and
(\ref{INDUNIFORM}) by Monte Carlo techniques. Whether this makes
sense near the physical number of dimensions, $d \to 4$, remains
to be seen. Clearly, one has to deal with the usual UV divergences
in this case.


\end{document}